\documentclass[prl,amsmath,amssymb,twocolumn,nofootinbib]{revtex4-2}
\usepackage{bm,color}
\usepackage[colorlinks=true,linkcolor=blue,urlcolor=blue,citecolor=blue]{hyperref}
\usepackage{graphicx}

\usepackage{times}

\definecolor{ashgrey}{rgb}{0.6, 0.65, 0.61}

\newcommand{\Eq}[1]{Eq.~(\ref{#1})}
\newcommand{\eq}[1]{(\ref{#1})}
\newcommand{\ds}[1]{\displaystyle }
\newcommand{\half}{\frac12}

\newcommand{\bea}{\begin{eqnarray}}
\newcommand{\eea}{\end{eqnarray}}
\newcommand{\beq}{\begin{equation}}
\newcommand{\eeq}{\end{equation}}

\newcommand{\rme}{\mathrm{e}}

\newcommand{\nn}{\nonumber}
\newcommand{\ca}[1]{{\cal #1}}
\newcommand{\be}{\begin{equation}}
\newcommand{\ee}{\end{equation}}
\newcommand{\Fig}[1]{\includegraphics[width=\columnwidth]{./#1}}  
\newcommand{\fig}[2]{\includegraphics[width=#1\columnwidth]{./#2}}

\newlength{\bilderlength}

\renewcommand{\section}[1]{{\it #1.}}

\begin{document}

\bibliographystyle{KAY-hyper}

\title{Hyperuniformity   in the Manna Model, Conserved Directed Percolation and Depinning}
\author{Kay J\"org Wiese}
  \affiliation{CNRS-Laboratoire de Physique de l’Ecole Normale Supérieure, PSL Research University, Sorbonne Université, Université Paris Cité, 24 rue Lhomond, 75005 Paris, France.}

\begin{abstract}
%\centerline{\today \ --\ \jobname.tex\ --\ compilation \input{./\jobname.counter}}

Hyperuniformity is an emergent property, whereby the structure factor   of the density $n$  scales as $S(q) \sim q^\alpha$, with $\alpha>0$.
We show that for the conserved directed percolation (CDP) class,   to which   the  Manna model belongs, there is an exact mapping between 
the density $n$ in CDP, and the interface position $u$ at depinning, $n(x)=n_0+\nabla^2 u(x)$, where $n_0$  is the conserved particle density. As a consequence, the hyperuniformity exponent  equals 
$\alpha={4-d-2\zeta}$, with  $\zeta$   the roughness exponent at depinning,  and $d$ the dimension. 
In   $d=1$,  $\alpha=1/2$, while  $0.6> \alpha\ge0$ for other $d$. Our results fit well simulations in the literature, except in $d=1$, where we perform our own   to confirm this result. 
Such an exact relation between two seemingly different fields is surprising, and paves  new paths to think about hyperuniformity and depinning. As  corollaries, we get results of unprecedented precision in all dimensions,  exact in $d=1$. This corrects   earlier work on   hyperuniformity  in CDP. 

\end{abstract}

\maketitle

\section{Context}
Hyperuniform (HU) structures have vanishing long-wavelength density fluctuations similar to   crystals, but no long-range order
\cite{TorquatoStillinger2003,Torquato2018,LeiNi2024}. 
The structure factor of the Fourier-transformed particle density, $S_q:= \left< n_q n_{-q}\right>$ vanishes for small $q$,  as $S_q\sim |q|^\alpha$ with $\alpha>0$. 
HU is observed in numerous systems \cite{Torquato2018,LeiNi2024}, ranging from   sandpile models \cite{BasuBasuBondyopadhyayMohantyHinrichsen2012,HexnerLevine2015,Lee2014,DickmanCunha2015,GrassbergerDharMohanty2016,Garcia-MillanPruessnerPickeringChristensen2018}, over sheared colloids  
\cite{WilkenGuerraLevineChaikin2021}, to densest packings \cite{WilkenGuoLevineChaikin2023}. 
All the above systems   have a critical state   recognized to be in the conserved directed percolation (CDP) class.
This attribution   usually relies on a comparison of   numerically measured critical exponents, especially the hyperuniformity exponnent $\alpha$. 
In this situation it is highly desirable to have an  analytical understanding of the underlying mechanism for hyperuniformity, 
and to know the relevant ritical exponents with precision. 
In this letter, we provide an exact mapping from CDP to depinning of an elastic manifold \cite{Wiese2021}. This mapping allows us to express the  hyperuniformity exponent $\alpha$ in terms of the dimension $d$ and the  roughness exponent $\zeta$ at depinning, 
\be\label{1}
\alpha = 4-d -2 \zeta.
\ee
Using $\zeta$ from for depinning gives $\alpha$ with   higher precision than  in most 
sandpile simulations, see Fig.~\ref{f:sigma}. 

This letter is organized as follows: we first review the concept of hyperuniformity, before introducing the   Manna sandpile,    the simplest and most prominent model in the  CDP  class. 
We then discuss further models in this class, and present the mapping. 
We finish with 
numerical evidence, and a discussion of relevant work in the literature.

\section{Hyperuniformity}
Consider a  particle system of size $L$, where the total number $N_{\rm tot}$ of particles is conserved.
We   ask how many particles $N_R$ are   in a part of the system of  radius $R\ll L$. If the system is translationally invariant, then 
\be
\left< N_R \right> = \frac{N_{\rm tot}}{L^d} R^d.
\ee
How does $N_R$ fluctuate? We expect that 
\be
\mbox{var}(N_R) = \left< N_R^2\right>  - \left< N_R\right> ^2 \sim R^{\kappa}.
\ee
One can show \cite{Beck1987} that (except for  fine-tuned models \cite{Beck2001})
\be\label{3}
d-1 \le \kappa \le d.
\ee
A Poisson process has   $\kappa = d$, a regular lattice $\kappa=d-1$.
When  $\kappa<d$ the system is said to be {\em hyperuniform}. This terminology was  introduced in  \cite{TorquatoStillinger2003} for $\kappa=d-1$, and  is now used for any  $\kappa<d$  \cite{DonevStillingerTorquato2005,ZacharyTorquato2009,GrassbergerDharMohanty2016}.
Alternatively, one can   consider the structure factor  of the Fourier transform $n_q$ of the density $n(x)$.
Its small-$q$ behavior is
\be\label{4}
S(q)= \left< n_q n_{-q}\right> \sim q^{\alpha}, \qquad \kappa+\alpha = d.
\ee
We are   interested in 
class-III HU systems \cite{Torquato2018}, which  correspond to  $0< \alpha\le1$.  Larger values of $\alpha$ are possible \cite{Torquato2018,LeiNi2019}, $\kappa$ then freezes at its lower bound $\kappa = d-1$.

\section{The Manna sandpile   and conserved directed percolation}
The Manna sandpile  \cite{Manna1991} is defined as follows: 
Consider a $d$-dimensional lattice, e.g.\ the checker board in $d=2$. Each site $x$ has $n(x)$ grains. 
If $n(x)\ge 2$,   with rate 1 move two of the grains, each  to a randomly chosen neighbor. 
This dynamics conserves the total number  $N:=\sum_x n(x)$ of particles.
Denote the fraction of $i$ times occupied sites as $a_i$.  Then (for each site $x$ and time $t$)  
$\sum_{i=0}^\infty a_i=1$, the number of particles is 
$\sum_{i=1}^\infty i\, a_i =n$, and the activity
$\sum_{i=2}^\infty (i-1)\, a_i = \rho$.
The last definition, introduced in \cite{Wiese2015}, gives a higher toppling rate to triple and higher occupied sites
than the standard definition. Since we are interested in densities  close to the transition, this does not matter \cite{Wiese2015}.
The benefit of this definition is the existence of the exact sum rule
\be\label{sum-rule}
n-\rho+e=1,
\ee
where $e:= a_0$ is  the fraction of empty sites.

The next step is to write effective stochastic equations of motion for $n$, $\rho$ and $e$. Due to the constraint \eq{sum-rule} 
there are two independent equations, usually  written in terms of particle density $n(x,t)$, and   activity $\rho(x,t)$ (for a derivation see \cite{Wiese2015}),
\begin{eqnarray}\label{eff:1}
\partial_t \rho(x,t) &=&     \nabla^2  \rho(x,t)+ \big[2n(x,t) {-}1\big] \rho(x,t)- 2 \rho(x,t)^2\nn\\
&&   + \sqrt{2\rho (x,t)}\,\eta(x,t) , \\
\partial_t n(x,t) &=&    \nabla^2  \rho(x,t) .  
\label{eff:2}
\end{eqnarray}
Here $\xi(x,t)$ is a standard white noise 
\be\label{5}
\left< \eta(x,t)\eta(x',t') \right> =   \delta^d(x-x')\delta(t-t').
\ee

\section{Sheared colloids}
The same effective model works for periodically  sheared colloids close to the reversible/irreversible transition. The connection can be understood via the 
Random Organization (RO) model \cite{WilkenGuerraLevineChaikin2021}: track  
the particle displacements after a full shear cycle of given amplitude. These displacements are   replaced by random ones of   observed amplitude,  for the active particles, i.e.\ those which     collided  
during the cycle. This results  again in the set of Eqs.~\eq{eff:1}-\eq{eff:2}.
The Biased Random Organization (BRO) model \cite{WilkenGuoLevineChaikin2023} is a variant, where colliding particles    receive an additional displacement moving them apart. 
In \cite{WilkenGuerraLevineChaikin2021,WilkenGuoLevineChaikin2023} the authors  claim that  RO and BRO both belong to the CDP class.  Furthermore,  BRO  is claimed to account for the statistics of random close packings (RCP) \cite{WilkenGuoLevineChaikin2023},
where other authors  claim RCP to be mean field in all dimensions \cite{CharbonneauKurchanParisiUrbaniZamponi2017}.

\section{Mapping CDP to depinning}
We now map   the CDP equations  \eq{eff:1}-\eq{eff:2} onto depinning. 
Instead of writing coupled equations for $n(x,t)$ and $\rho(x,t)$, use the  sum rule \eq{sum-rule} to   
 write coupled equations for $\rho(x,t)$ and $e(x,t)$,
 \begin{eqnarray}\label{233}
\partial_t e(x,t) &=&  [1{-} 2e(x,t)  ] \rho(x,t) + \sqrt{2\rho (x,t)}\,\eta(x,t) , \qquad  \\
\partial_t \rho (x,t) &=&    \nabla^2  \rho(x,t)  +\partial_t e(x,t). \label{234}  
\end{eqnarray}
To show the equivalence to disordered elastic manifolds   \cite{LeDoussalWiese2014a,JanssenStenull2016}, define
\bea\label{235}
\!\!\rho(x,t) &=\partial_t u(x,t) &\mbox{\quad (the velocity of the interface),}~~~~ \\
\!\!e(x,t)   &= {\cal F}(x,t) ~~~    &\mbox{\quad (the force acting on it).} 
\label{236}
\eea
Eq.~(\ref{234}) is  the time derivative of the equation of motion of an interface, subject to a random force ${\cal F}(x,t)$, 
\beq\label{230}
\partial_t u(x,t) =   \nabla^2 u(x,t) + {\cal F}(x,t).
\eeq
Remains to characterize the statistics of $\ca F$. 
Since $\rho(x,t)$ is positive for each $x$, $u(x,t)$ is  monotonously increasing. Instead of parameterizing ${\cal F}(x,t)$ by space $x$ and time $t$, it can be written as a function of space $x$ and {\em interface position} $u(x,t)$. Setting ${\cal F}(x,t) \to F\big(x,u(x,t)\big)$, \Eq{233}  becomes
\bea
\partial_t {\cal F}(x,t) &\to& \partial_t F\big(x,u(x,t)\big) \nn\\
&=& \partial_u F\big(x,u(u,t)\big) \partial_t u(x,t)  \nn \\
&=& \Big[1-2 F\big(x,u(x,t)\big)\Big] \partial_t u(x,t)  \nn\\
&&+ \sqrt{2 \partial_t u(x,t)} \eta(x,t).
\eea
For each $x$, this   is equivalent to an Ornstein-Uhlenbeck \cite{UhlenbeckOrnstein1930} process   $F(x,u)$, defined by  
\bea\label{OU}
\partial_u F(x,u) =  1-2 F(x,u) + \sqrt2\; \xi(x,u), \\
\left< \xi(x,u) \xi(x',u') \right> = \delta^d(x-x') \delta(u-u').
\eea
While the noise $\eta(x,t)$ is uncorrelated in time, $\xi(x,u)$ is uncorrelated in the interface position $u$. 
Given $x$, $F(x,u)$ is a Gaussian Markovian process with mean $\left< {F(x,u)}\right> =1/2$, and variance in the steady state of 
\be\label{cor:FF}
\left<{ \left[ F(x,u)-{\textstyle \half} \right] \left[ F(x',u')- {\textstyle \half} \right]}\right>= \frac12 \delta^d(x-x') \rme^{-2 |u-u'|}
.
\ee%
\begin{figure}[t]
%\fboxsep0mm
%\unitlength1mm
%{%\blue\fbox
%{\black
%\begin{picture}(86.5,54.5)
%\put(0,0){\Fig{alpha}}
%\put(10,11)
%{\begin{tabular}{|c|c|c|c|c|c|}
%\hline
%$d$ & 0 & 1 & 2& 3 & $\ge 4$ \\
%\hline 
%$\alpha^{\rm num}$ &$0$ & $0.5$ & $0.494(4)$ $ $ & $0.29(2)$ & $0$\\
%\hline
%$\alpha^{\rm FT}$ & $0$ & $0.5$ & $0.4964$ & $0.2868$ & $0$\\
%\hline
%\end{tabular}}
%\end{picture}}}
\Fig{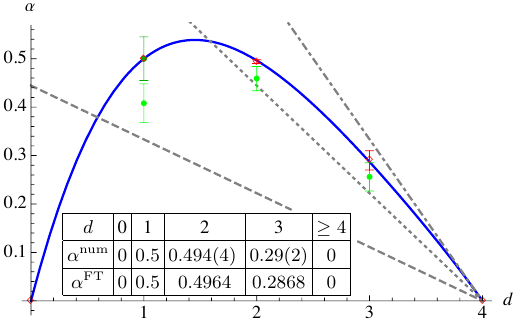}
\caption{The exponent $\alpha$ of the structure factor $S(q)\sim |q|^\alpha$ as a function of dimension $d$ for the  Manna model. Blue solid line from the $\epsilon$-expansion of \cite{SemeinkinWiese2024}, red dots (with error bars) simulations at depinning \cite{RossoHartmannKrauth2002,ShapiraWiese2023}.
Simulations in green are from \cite{HenkelHinrichsenLubeck2008}. The dark green data point is from Fig.~\ref{Sqq}. In gray are the different $\epsilon$-expansion results, $\alpha=\epsilon/9$ (dashed) \cite{HexnerLevine2015}, $\alpha=2\epsilon/9$ (dotted)\cite{MaPauschCates2023} and $\alpha=\epsilon/3$ (dot-dashed) (leading term of \Eq{22}).}
\label{f:sigma}
\end{figure}% 
Writing the equation of motion (\ref{230}) as  
\beq 
\partial_t u(x,t) =  \nabla^2 u(x,t) + F\big(x,u(x,t)\big), 
\eeq
it is   the equation of motion of an interface with {\em position} $u(x,t)$, subject to a {\em quenched disorder force} $F\big(x,u(x,t)\big)$. The latter is $\delta$-correlated in the $x$-direction, and short-ranged correlated in the $u$-direction: it  is a disordered elastic manifold subject to Random-Field (RF) disorder. 
As a consequence, results for  disordered elastic manifolds can   be used for CDP and the Manna model. 

\section{Hyperuniformity in the Manna model}
\begin{figure*}[t]
\fig{0.6}{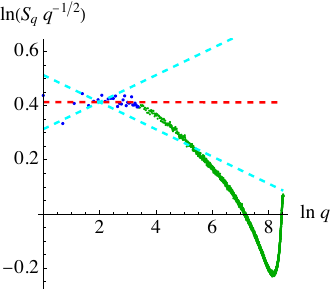}\hfill\fig{0.6}{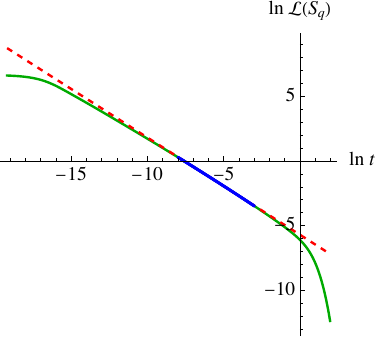}\hfill\fig{0.6}{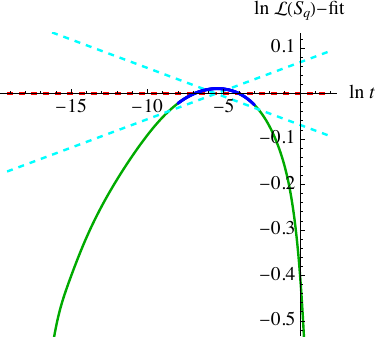}
\caption{Left: the compensated structure factor $S_q q^{-1/2}$ in a ln-ln plot for a periodic system of size $L=10^5$, with $8\times 10^7$ samples. The red dashed line with slope $0$ indicates the behavior $S_q\sim \sqrt q$, the cyan curves   power laws with an exponent deviating by $\pm 0.05$, indicating our interval of confidence.  The middle plot shows the Laplace-transform $\ca L_{\beta=1}\circ S(t)$ defined in \Eq{Laplace-transform}. The fit   by the red dashed line has slope $-(1+\alpha)=-3/2$. Subtracting this fit gives the plot shown to the right.}
\label{Sqq}
\end{figure*}%
The roughness exponent $\zeta$ for the random manifold is defined via 
\be
\left < [u(x,t)-u(y,t)]^2 \right> \sim |x-y|^{2\zeta}. 
\ee
\Eq{235}   implies that $\rho$ is not HU, 
\be\label{21}
\left< \rho(x,t) \rho(y,t) \right>^{\rm c} \sim |x-y|^{2(\zeta-z)},
\ee
where $z$ is the dynamical critical exponent \cite{Wiese2021}.
As a new result, let us calculate the particle-density correlation function. We   have to identify $n(x,t)$ with the appropriate random-manifold field. Using Eqs.~\eq{eff:2} and   \eq{235}, we find $\partial_t n(x,t)=\nabla^2 \partial_t u(x,t)$, or after integration over time
\be\label{19}
n(x,t) = \nabla^2 u(x,t) +n_0.
\ee
Here $n_0$ is the conserved mean density of particles, i.e.\ the  conserved total number of particles divided by the volume.
Taking the derivatives implied by  \Eq{19} yields
\be
\left< n(x,t) n(y,t)  \right>^{\rm c} \sim |x-y|^{2\zeta-4}. 
\ee
In Fourier space this implies our result  \eq{1}, 
\be\label{prediction-sigma}
S_q := \left< n_q n_{-q}\right> \sim |q|^{\alpha}, \quad \alpha=4-d-2\zeta. 
\ee
Denoting $\epsilon = 4-d$,  and using for $\zeta$ its  $\epsilon $ expansion $\zeta=\frac \epsilon3 + \zeta_2 \epsilon^2 + \zeta_3 \epsilon^3$, see \cite{ChauveLeDoussalWiese2000a,LeDoussalWieseChauve2002} (2-loop) and \cite{SemeinkinWiese2024} (3-loop), $\alpha$ becomes
\bea\label{22}
\alpha &=& \epsilon - 2 \zeta = \frac\epsilon3 -2\zeta_2 \epsilon^2 - 2 \zeta_3 \epsilon^3 + \ca O(\epsilon^4)\\
\zeta_2 &=&  0.0477709715468230578...\\
\zeta_3 &=& -0.0683544 (2).
\eea
To obtain predictions for $\alpha$ in the CDP class, we can  
use \Eq{22} via Padé-Borel resummation supplemented 
by the knowledge of $\zeta_{d=0}=2$ \cite{Wiese2021}, and $\zeta_{d=1}=5/4$ 
\cite{GrassbergerDharMohanty2016,ShapiraWiese2023}. This leads to 
\be
\alpha_{d=1}^{\rm FT} = 1/2  ,\quad  \alpha_{d=2}^{\rm FT}=0.4964, \quad   \alpha_{d=3}^{\rm FT} = 0.2868 .
\ee 
Alternatively, use the  best   simulation results   
$\zeta_{d=2}= 0.753\pm 0.002$ \cite{RossoHartmannKrauth2002} and $\zeta_{d=3} = 0.355\pm 0.01$ \cite{RossoHartmannKrauth2002}, 
to find   
\be
\alpha_{d-2}^{\rm num} = 
 0.494 (4), \quad 
 \alpha_{d=3}^{\rm num} = 0.29(2).
\ee
As Fig.~\ref{f:sigma} shows,  $0\le \alpha<1 $ in all dimensions, the signature given in Eqs.~\eq{3}-\eq{4} for a class-III hyperuniform system. 
The figure compares $\epsilon$-expansion, numerical simulations for $\alpha$ \cite{HenkelHinrichsenLubeck2008} in the Manna model  (see below), and predictions using \Eq{prediction-sigma} with $\zeta$ from simulations at depinning.

\section{Active state}
When disordered elastic manifolds are driven at a finite velocity $v$, the force correlations 
become $\delta$-correlated in time \cite{Wiese2021}, and act like a thermal noise, leading to a  roughness exponent   $\zeta_{\rm moving} =\frac{2-d}2$.  
This   gives  the hyper\-uniformity exponent in the active phase, 
\be
\alpha_{\rm active} = 4-d- 2\zeta_{\rm moving} =  2. 
\ee
This was   observed in the active phase of the RO and Manna models with center of mass conservation \cite{HexnerLevine2017}, as well as in non-equilibrium hyperuniform fluids \cite{LeiNi2019}.

\section{Stability of CDP, and relation to DP}
There was a long debate whether the Manna model, or the corresponding CDP theory, are in the same universality class as 
disordered elastic manifolds or whether they   belong to a different universality class,    the directed-percolation (DP) class.
This question was finally settled in  \cite{LeDoussalWiese2014a} by the arguments presented above. 
To understand how robust CDP is, replace in  \Eq{eff:1} the term $[2n-1]\rho\to \lambda [2n-1]\rho$, while keeping $n$ as a (possibly unobservable) variable.
The limit of $\lambda \to 0$ corresponds to  DP.  
 This changes \Eq{OU} for $F(x,u)$ to 
\be
\partial_u F(x,u) =  \lambda[1 -2 F(x,u)] -2 (1-\lambda) \rho + \sqrt2\; \xi(x,u).
\ee
Compared to \Eq{OU}, 
it  has an additional noise proportional to $\rho$, with both a mean and a variance. We expect that 
 for given $x$, as long as $\lambda >0$, the process $F(x,u)$ remains short-range correlated with a correlation length   $\xi_{F} \approx 1/\lambda$. (This conclusion was reached via a different argument in  \cite{LeDoussalWiese2014a}.)  While the correlation length $\xi_F$  diverges  for $\lambda \to \infty$, we expect the CDP class to be robust as long as $\lambda>0$, i.e.\ as long as there is a conserved density $n$, and it appears via a term proportional to $n\rho$ in the equation for $\partial_t \rho$. 
It would be interesting to repeat simulations on sheared colloids \cite{TjhungBerthier2015}, for which  opposite conclusions were reached.

\section{Improved numerical checks}
There is some tension between simulation results   $\alpha^{\rm Manna}_{d=1} =0.41(4)$ \cite{HenkelHinrichsenLubeck2008},  
the seemingly accepted value 
  $\alpha^{\rm RO}_{d=1}\approx 0.45$ \cite{WilkenGuerraLevineChaikin2021,WilkenGuoLevineChaikin2023,LeiNi2024}, and our exact result $\alpha^{\rm exact}_{d=1}=1/2$.
For this reason we 
performed numerical simulations for Manna with systems of size up to $L=10^4$. The results of the latter compensated for the predicted behavior are shown on the left of Fig.~\ref{Sqq}.
There are   strong finite-size corrections which make understandable the relatively small value given in \cite{HenkelHinrichsenLubeck2008}.
However, in the relevant limit of small $q$, the data are consistent with $\alpha=0.5$ (red dashed line), while 
the cyan (bright)  lines for  $\alpha=0.45$ and $\alpha=0.55$ are the confidence interval reported on Fig.~\ref{f:sigma}.
To reduce the statistical noise, we also show the results for a generalized Laplace-transform,
\be\label{Laplace-transform}
\ca L_\beta \circ S(t) :=\sum_q \rme^{-|q|^\beta t} S_q \sim   t^{-\frac{1+\alpha}{\beta}}.
\ee
The value $\beta=1$ is the standard Laplace-transform, and was used e.g.\ in \cite{RossoLeDoussalWiese2009a}. $\beta=2$ is now popular under the name ``diffusion spreadability'' \cite{Torquato2021}.  Our data analysis shows $\beta=1$, $2$ or $4$ to be equivalent for all practical purposes.  As Fig.~\ref{Sqq} for $\beta=2$ reveals, the noise is indeed reduced, but it is more difficult to choose 
the proper domain to fit to. All   fits give   $\alpha = 0.5\pm 0.05$. 

\begin{figure}
\fig{0.5}{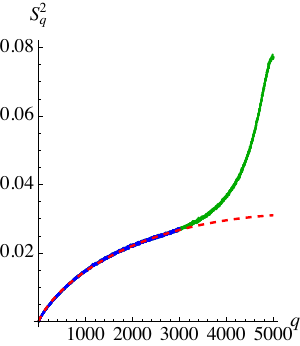}\fig{0.5}{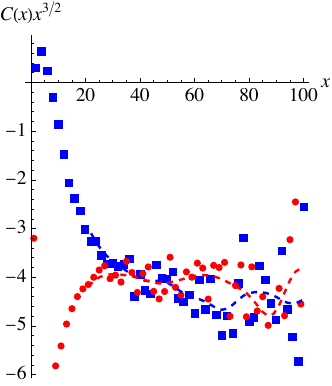}
\caption{Left: fit (red dashed) of $S_q^2$ (solid blue used for fit,   green not used) for $L=10^5$ with   $S_q^2 \simeq 5.31 \cdot 10^{-6}q \ln(16111/q)$. Right: The compensated correlation function $C(x)x^{3/2}$ for even  (blue squares) and odd (red discs) distances $x$. In dashed weakly filtered data as guide for the eye. One sees   strong even/odd lattice effects, which start to disappear at $x\approx 30$.}
\label{f:Sqq-with-log}
\end{figure}

The reader may wonder where this problem in such a large system comes from, and whether there might  be systematic corrections. 
While there is   no proper theoretical motivation, on a phenomenological level  the deviations from a pure power law are well   fitted with a  logarithm, as  Fig.~\ref{f:Sqq-with-log} attests.
To proceed, it is 
instructive to plot the density correlations as a function of distance. 
For short even distances, we find positive correlations, due to events where one grain is moved to the right, and one   to the left. These positive correlations become negative for $x\ge 8$, but one has to wait to $x\approx 30$ until even and odd correlations are comparable. This indicates that $\ell=30$ is the minimal coarse graining size, taking out $1.5$ decades from the fitting range for $S_q$, certainly one reason for its  slow convergence.
One may also wonder whether this is related to the saturation of the apparent roughness exponent at depinning $\zeta_{\rm app}^{\rm dep}(d=1)\approx 1$ \cite{LeschhornTang1993,Wiese2021}.

\section{Relation to the literature}
Our results contradict two works from the literature:  the phenomenological observation   $\alpha = \epsilon/9$ \cite{HexnerLevine2015} (we  supplemented \cite{HexnerLevine2015} with $[\rho]_L = \zeta-z$ as  implied by \Eq{21}) and $\alpha = 2\epsilon/9$   \cite{MaPauschCates2023} obtained from RG within the Doi-Peliti approach.
None of these works uses functional RG, which is  crucial to account for the non-trivial structure present at 2-loop  \cite{ChauveLeDoussalWiese2000a,LeDoussalWieseChauve2002} and 3-loop  \cite{SemeinkinWiese2024} order
at depinning. Ref.~\cite{MaPauschCates2023} does this calculation in terms of active and passive particles in a 2-species model. The density of the latter 
is a linear combination of fields used here, 
$n_{\rm p} =a_1 \approx n-2\rho$. Since  $n -n_0= \nabla^2 u$ and $\rho = \partial_t u$, the  scaling dimensions of the two terms differ by $z-2=\ca O(\epsilon)$. As a result,  $n_{\rm p}$ is not a proper scaling field of the RG, a problem   known in other contexts \cite{KavirajRychkovTrevisani2022}. As the two fields are degenerate at $\epsilon=0$, their respective $\ca O(\epsilon)$  corrections are easily attributed to the $\ca O(\epsilon)$ correction of their linear combination $n_{\rm P}$. 

A  new feature of   \cite{MaPauschCates2023} is the introduction of a  current noise in their Eqs.~(4) and (5). As additional terms have to vanish in the absorbing state $\rho=0$, 
we may add to \Eq{233} the divergence of a current; the most relevant is   $\nabla \vec J(x,t)\equiv \nabla \partial_t \vec \Gamma(x,t)$, with  $\partial_t \vec\Gamma(x,t) = \sqrt{\rho(x,t)}  \vec\eta(x,t)$,   $
\left<\eta^i(x,t)\eta^j(x',t')\right> = \delta^d(x-x')\delta(t-t')\delta^{ij}$. Comparing $\nabla^2 \rho \sim L^{\zeta-z-2}$ to $ \nabla \sqrt{\rho(x,t)} \vec\eta(x,t)\sim L^{\frac{\zeta-z}2-1-\frac{d+z}2}$ we conclude that the latter is perturbativly irrelevant as long as $d+\zeta>2$, which is satisfied for all $d>0$.
It may become relevant non-perturbatively: using the same techniques as in the derivation of \Eq{cor:FF},  $\Gamma^i(x,t)$ has the statistics of a random walk, $ \langle  [ \Gamma^i(x,t)-\Gamma^i(x,0) ]^2\rangle \sim |u(x,t){-}u(x,0)|$. Using results of the Brownian-force model  \cite{Wiese2021} one gets $\zeta= \zeta_{\rm BFM}-1 =3-d$. This would destroy HU in dimension $d=1$, in contradiction to   simulations. As the current in the Manna model should not keep an infinite-time memory as a random walk does,  we propose to modify the equation for $\vec\Gamma$ to an Ornstein-Uhlenbeck process in $u$, as in Eqs.~\eq{233} and \eq{OU}:  $\partial_t \vec\Gamma(x,t) = \sqrt{\rho(x,t)}  \vec\eta(x,t) - \kappa \rho (x,t)\vec \Gamma(x,t)$, $\kappa>0$. This takes into account that if two particles jump onto the same neighbor, contributing to the current $J$, 
 a toppling will take place there, resulting (possibly after   iteration) into a counter-current. $\Gamma^i(x,t)$   then has correlations as \Eq{cor:FF}, and $\nabla \partial_t \vec \Gamma(x,t)$ is irrelevant. 
While the mapping from the CDP equations  \eq{eff:1}-\eq{5}  to depinning  is exact, the additional irrelevant 
current in Manna makes interfaces constructed as $u(x,t):=\sum_{\tau=0}^t    \rho(x,\tau)$ microscopically rough. This can be  seen in simulations.

\section{Conclusions} 
In this letter, we have shown how hyperuniformity in   CDP   is related to depinning. This equivalence yields   
precise theoretical predictions for the hyperuniformity exponent in all dimensions, both close to the transition, and in the active state.  It would be interesting to extent these results to other universality classes, as qKPZ \cite{TangKardarDhar1995,RossoHartmannKrauth2002,MukerjeeBonachelaMunozWiese2022,MukerjeeWiese2022}, and to see whether the fascinating phenomenology in plastic depinning \cite{ReichhardtReichhardt2016} has an equivalence in sandpile models. 

\begin{acknowledgements}
I thank Duyu Chen and Ran Ni for sharing their knowledge on hyperuniform systems, and X.~Ma, J.~Pausch  and M.E.~Cates for stimulating discussions.
\end{acknowledgements}
%
%
%\bibliography{../../../citation/citation}

\ifx\doi\undefined
\providecommand{\doi}[2]{\href{http://dx.doi.org/#1}{#2}}
\else
\renewcommand{\doi}[2]{\href{http://dx.doi.org/#1}{#2}}
\fi
\providecommand{\link}[2]{\href{#1}{#2}}
\providecommand{\arxiv}[1]{\href{http://arxiv.org/abs/#1}{#1}}
\providecommand{\hal}[1]{\href{https://hal.archives-ouvertes.fr/hal-#1}{hal-#1}}
\providecommand{\mrnumber}[1]{\href{https://mathscinet.ams.org/mathscinet/search/publdoc.html?pg1=MR&s1=#1&loc=fromreflist}{MR#1}}

\end{document}